\documentstyle[aps,prl,amsfonts,amsbsy]{revtex}
\twocolumn
\draft
\begin{document}
\title{Comment on `Renormalization-Group Calculation of the Dependence on
Gravity
of the Surface Tension and Bending Rigidity of a Fluid Interface'}
\maketitle
In a recent Letter \cite{SR01} a renormalization-group (RG) calculation of the
gravity dependence of properties of a  liquid-vapor interface is
presented whose central point is the use of the
(bare) capillary-wave-like Hamiltonian
\begin{equation}
{\cal H}_{CW}[\zeta ]={\int }\!d^{d-1}R\,{\Big[ {\frac{\gamma _0}2}\,{%
\left| \nabla \zeta \right| }^2+{a_0(g)\over 2}\,\zeta ^2+\frac \lambda {4!}%
\,\zeta ^4\Big] }  \label{Ham}
\end{equation}
to study the interface in the limit of vanishing gravitational acceleration $%
g$. Here $\zeta (\vec{R})$ is the local height of the interface
configuration $[\zeta ]$ at position $\vec{R}$ relative to the
plane $z=0$. Recognizing that (\ref{Ham}) is just the usual $\varphi ^4$
Hamiltonian and  that $a_0(g)$ vanishes as $g\rightarrow 0$, the
authors followed the standard RG analysis for the
critical {\em Ising} system,  arrived at the fixed point Hamiltonian
\begin{equation}
{\int }\!d^{d-1}R\,{\Big[ {1\over 2}\,{| \nabla \zeta | }^2+%
\frac {\lambda ^*}{4!}\,\zeta ^4\Big] }\;, \label{fpH}
\end{equation}
and reached  a remarkable conclusion: interfacial properties (in $d$
dimensions) for $g=0$ should be identical to those in a critical,
 $d-1$ dimensional bulk Ising system.

In order for this analysis and its conclusions to be valid,
the (renormalized) coupling constant  $\lambda $ {\em must remain positive}
in this limit, i.e., $\lim_{g\to 0}\lambda (g)=\lambda ^{(0)}>0$. However,
in the present interface case (in contrast to that of critical bulk systems),
there are
fundamental symmetry requirements  that
enforce $\lambda ^{(0)}=0$.
We believe that the authors' introduction of a nontrivial $\lambda (g)$
is justifiable. But as we shall show in the sequel,
their assumption of $\lambda ^{(0)}>0$ is untenable, so that
their  subsequent analysis is irrelevant and the conclusion unreliable.

Consider a simple liquid-vapor system in $d$ dimensions, under the influence
of a
%gravitational
potential $g\,z$. Let $(x_i)=(R_\alpha ,x_d\equiv z)\in {%
\mathbb{R}}^d$
%, $\alpha =1,\ldots ,d{-}1$,
 be the coordinates of a point
relative to a fixed frame with Euclidean unit vectors $\vec{e}_i$. In the
{\em absenc}e of gravity, the system respects the Euclidean symmetries and
hence is invariant under translations and
rotations $x_i\to x_i^{\prime }={{\cal R}_{ij}}\,x_j-A_i$, where $({\cal R}%
_{ij})$ is a rotation matrix. In the two-phase region (of the $g=0$
system), these symmetries are broken spontaneously by the presence of an
interface, so that {\em nonlinear} realizations of the Euclidean group are
necessary. For $g\ne 0$, these symmetries are  broken
explicitly. Under a rotation, the system transforms into one with a rotated
field: $g\,\vec{e}_d \to  {\mathcal{R}}_{id}\,g\,\vec{e}_i$.

Let us write possible (bare) interface Hamiltonians
as
\begin{equation}
{\cal H}^{(g)}[\zeta ]={\int }\!d^{d-1}R\,{\cal L}^{(g)}(\zeta ,\partial
\zeta ,\partial ^2\zeta ,\ldots )\;,
\end{equation}
where $\partial \zeta $, $\partial ^2\zeta $,\ldots stand for first,
second, and higher partial derivatives $\partial \zeta /\partial R_\alpha$
etc.
${\cal H}^{(0)}[\zeta ]$ {\em must be }%
invariant under the transformations discussed above. We consider separately
the consequences of this trivial observation.

{\it (i) Translations along  the $z$ axis:}  This symmetry is realized by
$\zeta \rightarrow \zeta ^{\prime }=\zeta -A_d$. The invariance of
${\mathcal{H}}^{(0)}$ means that the energy cost of such a translation vanishes as $g\to 0$. Hence ${\cal L}^{(0)}$ cannot depend explicitly on $\zeta $: we must have
${\cal L}^{(g)}\to {\cal L}^{(0)}(\partial \zeta ,\partial ^2\zeta ,\ldots )$.
 This  rules out
terms like $\lambda\, \zeta ^4$, as well as (\ref{fpH}), in the limit $g\to 0$.
Yet, $\lambda ^{(0)}>0$ is crucial for the approach in \cite
{SR01}, the presence of an upper critical dimension of $5$, and the fixed
point (\ref{fpH}).

{\it (ii) Rotations about $\vec{e}_\alpha $:} As discussed in Ref.\ \cite
{WZ79}, ${\cal H}^{(0)}$ must realize this symmetry in a
nonlinear fashion. Specifically, if only first derivatives $\partial \zeta $
are taken into account, then ${\cal H}^{(0)}$ is simply proportional to the
surface area, i.e.,
\begin{equation}
{\mathcal{L} }^{(0)}(\partial \zeta ,0,\ldots )=\sigma \,\sqrt{%
\eta }
\;,\label{drumhead}
\end{equation}
where $\eta =1+(\partial _\alpha \zeta )^2$ is the determinant of the
induced metric. In fact, this `drumhead' model can be derived directly
from appropriate  bulk models of the $\varphi ^4$ type
in the low-temperature limit \cite{DKW80}. Corrections to it
depending on derivatives of higher than
first order
%(which must also be reparametrization invariant)
 can be derived
in a systematic fashion \cite{Zia85,LL83}; the leading ones involve the
trace of the  curvature tensor and the curvature scalar.

Starting from a $\varphi ^4$ model for the {\em bulk}, the parameters $a_0$
and $\lambda $ could be computed by an appropriate generalization of the
analysis of Refs.\ \cite{DKW80} and \cite{Zia85} to the case $g\ne 0$. That
these parameters vanish for $g=0$ has been shown already in these references.

In summary, the interface model of Ref.~\cite{SR01}, while valid
for $g\ne 0$, violates fundamental symmetry requirements for $g=0$.
Thus its results cannot be applied to critical properties of
interfaces in vanishing gravitational fields.

One of us (H.W.D.) is indebted to S.\ Dietrich for calling his attention to
Ref.~\cite{SR01}. \\[0.5em]
 H.~W. Diehl,  Fachbereich Physik, Universit{\"a}t Essen,
D-45117 Essen, Germany\\[0.5em]
R.~K.~P. Zia,
 Department of Physics,
 Virginia Polytechnic Institute and State University,
Blacksburg, Virginia 24601, USA

\end{document}